\documentclass[12pt,preprint]{aastex}

\def\lsim{~\raise0.3ex\hbox{$<$}\kern-0.75em{\lower0.65ex\hbox{$\sim$}}~}
\def\gsim{~\raise0.3ex\hbox{$>$}\kern-0.75em{\lower0.65ex\hbox{$\sim$}}~}

\def\gt{~\hbox{$>$}~}
\def\msun{M_\odot}
\def\lbrack2{[\![}
\def\rbrack2{]\!]}

\def\tdyn{t_{\rm dyn}}
\def\tcool{t_{\rm cool}}
\def\1D{one spatial dimension}
\def\2D{two spatial dimensions}
\def\3D{three spatial dimensions}

\def\nh{{n_{\rm H}}}

\def\i0g{{I_{0\rm g}}}
\def\e0g{{E_{0\rm g}}}

\def\hp{h_{\rm P}}

\def\kms{{\rm\,km\,s^{-1}}}
\def\kpc{{\rm\,kpc}}
\def\mpc{{\rm\,Mpc}}
\def\s{{\rm\,s}}

\def\msun{{\rm\,M_\odot}}

\def\cm{{\rm\,cm}}
\def\erg{{\rm\,erg}}
\def\ster{{\rm\,sr}}
\def\hz{{\rm\,Hz}}

\def\te{{T_{\rm e}}}

\def\kb{{k_{\rm B}}}

\def\tento#1{\times 10^{#1}}

\newcommand\hi{\hbox{H\,{\sc i}~}}
\newcommand\hii{\hbox{H\,{\sc ii}~}}

\def\euv{{\epsilon_{\rm UV}}}
\def\fesc{{f_{\rm esc}}}

\slugcomment{}

\shorttitle{Numerical Radiative Transfer}
\shortauthors{Razoumov et al.}

\begin{document}


\title{Cosmological Hydrogen Reionization with Three Dimensional
Radiative Transfer}


\author{Alexei O. Razoumov and Michael L. Norman}
\affil{Center for Astrophysics and Space Sciences,
University of California, San Diego, CA 92093}

\author{Tom Abel}
\affil{Harvard Smithsonian Center for Astrophysics}

\author{Douglas Scott}
\affil{Department of Physics and Astronomy,
University of British Columbia, Vancouver, Canada V6T 1Z1}
\email{aastex-help@aas.org}




\begin{abstract}
We present new calculations of the inhomogeneous process of
cosmological reionization by following the radiative transfer
carefully in pre-computed hydrodynamical simulations of galaxy
formation. These new computations represent an important step on the
way towards fully self-consistent and adaptive calculations which will
eventually cover the enormous range of scales from sizes of individual
mini-halos to the mean free path of ionizing photons in the
post-overlap Universe.  The goal of such simulations is to include
enough realistic physics to accurately model the formation of early
structures and the end of the `dark ages'.  Our new calculations
demonstrate that the process by which the ionized regions percolate
the Universe is complex, and that the idea of voids being ionized
before overdense regions is too simplistic.  It seems that
observational information pertaining to the reionization epoch may now
be in our grasp, through the detection of Gunn-Peterson troughs at
$z\sim6$.  If so, then the comparison of information from many lines
of sight with simulations such as ours may allow us to disentangle
details of the ionization history and trace the early formation of
structure.
\end{abstract}




\section{Introduction}

Including radiative transfer (RT) into three dimensional simulations
of the first structures and galaxy formation has proven to be
extremely challenging.  This is mostly because of the non-locality of
radiation physics.  Although many different aspects of RT can be
accounted for by various approximations (e.g., space-averaged field,
self-shielding, diffusion approximation), so far most calculations
have not been able to include details of the spatially inhomogeneous
3D transfer of radiation.  Of particular interest is studying the
effects of radiative feedback from the first luminous structures in
the universe, which is the focus of the present study.  Photons from
these early objects heated and ionized the intergalactic medium (IGM),
altering the dynamics of baryons on a wide range of scales.

Intriguingly, the process through which the ionized regions percolated
the Universe may be directly observable in the spectra of the most
distant quasars \citep{becker01,djorgovski01}. In order to use such
observations to understand early structure formation it is important
to model the relevant physics as reliably as possible.  For this
purpose accurate treatment of the radiation is crucial.

The RT equation in the expanding Universe is given by

\begin{equation}
{1\over c}{\partial I_\nu\over\partial t}+{{\bf n}\over
a}\cdot{\bf\nabla}I_\nu- {H\over c}\left(\nu{\partial
I_\nu\over\partial\nu}-3I_\nu\right)= \epsilon_\nu-\kappa_\nu I_\nu ,
\end{equation}

\noindent
which is the conservation law for the specific intensity $I_\nu$
propagating in the direction ${\bf n}$, with $H$ being the expansion
rate, $a=1/(1+z)$ the cosmological scale factor and $\epsilon_\nu$ and
$\kappa_\nu$ the emission and absorption coefficients,
respectively. Unfortunately, full solution of this seven dimensional
(three in space, two angles, frequency and time) equation is still
well beyond our computational capabilities. Although in many
astrophysical situations it is possible to reduce the dimensionality
of this equation, transfer in the clumpy IGM has proved to be one of
the most difficult problems, since it does not provide us with any
obvious spatial symmetries. For this reason, a great deal of effort
has been put into solution of the spatially and directionally averaged
RT equation \citep{haardt96,chiu00,valageas99}. In these models the UV
background is computed using the properties of an average cosmological
volume, such as the mean luminosity function per unit volume for the
source term and the statistical properties of absorbing clouds (often
in the form of the clumping factor) for the recombination term. This
part of the calculation is sufficiently simple that it is then
possible to estimate in great detail the spectrum of the background
radiation (Haardt \& Madau 2001). However, these models fail to
address questions related to the time-dependent propagation of
radiation fronts and therefore cannot predict what fraction of this
radiation leaves host halos or how this radiation is being deposited
into the IGM.

In the past few years, a number of different techniques for practical
solution of the 3D RT equation have been suggested. Among the first
fully numerical works, Umemura, Nakamoto, \& Susa (1998) calculated
reionization of a cosmological volume by a uniform UV background from
$z\,{=}\,9$ to $z\,{=}\,4$, solving the full 3D quasi-static RT
equation on a massively parallel architecture at very high numerical
resolution, $128^3\times 128^2$ (spatial $\times$ angular), using the
method of short characteristics. However, their calculations show that
direct integration at full angular resolution is very costly,
requiring as much as $\sim 100$ hours to reconstruct a single snapshot
on 256 processors. Abel, Norman, \& Madau (1999) developed a ray
tracing algorithm for radial RT around point sources which conserves
energy explicitly, and thus gives the right speed of Ionization fronts
(hereafter I-fronts).  However, in its original form this algorithm
will work only for a small number of sources, since its operation
count goes as ${\cal O}(N\times N_{\rm src})$, where $N$ is the number
of grid cells or particles in the volume, and $N_{\rm src}$ is the
number of sources. \citet{abel01} have modified this algorithm,
introducing trees of ray segments which recursively split into
sub-segments as one goes farther away from the source, resulting in a
significant speed up of calculations. \citet{razoumov99} have
developed a different technique which is essentially a poor man's
solution to the 5D (three spatial coordinates and two angles)
advection equation designed to work for both point sources and the
diffuse flux. Unfortunately, it requires one to store the 5D
intensity, hence, this technique is limited to very low angular
resolution making it currently impractical for high-resolution
cosmological simulations. Recently, \citet{ciardi01} implemented a
fast Monte Carlo RT method to study propagation of I-fronts around a
proto-galaxy at $z=12$ showing that reasonable integration times are
possible for a single source.  Further numerical and analytical
studies of reionization are discussed in a comprehensive review by
Loeb \& Barkana (2001).

Since all of these techniques try to use a fair sampling of the
multidimensional phase space to directly solve the 3D RT problem, they
tend to take at least ${\cal O}(N^2)$ operations. Two
notable exceptions are the local optical depth approximation (Gnedin
\& Ostriker 1997), requiring ${\cal O} (N)$ operations, and the explicit
moment solver {\sl OTVET} (optically thin variable Eddington tensor
formalism) of Gnedin \& Abel (2001), reducing the operation count to
${\cal O}(N\log N)$. The local optical depth approximation, however,
does not calculate the optical depth between two points (the emitter
and the point where the radiation field is to be computed), replacing
it instead with local quantities, and thus making it possible to
compute the radiation field with a fast gravity solver, in this
drastic approximation at least.  The explicit
moment solver (Gnedin \& Abel 2001) retains the time derivative of the
RT equation and solves hyperbolic (advection) moment equations,
assuming that the geometry of the radiation field (given by the
components of the Eddington tensor) is the same as in the optically
thin regime.  The assumption of optically thin Eddington tensors,
although exact in the limit of a single point source, may break down in
complex situations with very inhomogeneous source functions and opacity
fields.

In the following we use yet another approach, combining a photon
conservation algorithm with two independent hierarchies of trees, one
for the rays \citep{abel01} and one for the stellar sources. This
approach is similar to the work of Sokasian, Abel, \& Hernquist (2001)
who computed helium reionization for a large number of short-lived
sources using the radial ray technique of Abel et al.~(1999) to
integrate the ionization jump condition (Abel 2000).  In a similar way
we also follow the RT in a pre-computed density field. However, we
cover a different redshift interval and follow the full time-dependent
radiative transfer.

In the next section we describe the galaxy formation simulations used
as input for the solution of the radiative transfer. Then we describe
in detail our method for radiative transfer. Finally we discuss the
properties of the inhomogeneously ionized intergalactic medium and the
radiation field derived from the radiative transfer simulations. 

\section{Input Simulations}

Tassis et al.~(private communication) have performed cosmological
three dimensional simulations using the adaptive mesh refinement (AMR)
code {\sl enzo} of Bryan \& Norman (1999), including star formation
and feedback. We give only a brief summary of the simulations
parameters here, since a detailed analysis of the results will be
given elsewhere.  The simulations follow the dark matter and
hydrodynamics of the baryons starting at a redshift of 60. The initial
conditions were set up using the power-spectra for cold dark matter
and gas computed by CMBfast (Seljak \& Zaldarriaga 1996) appropriate
for a flat cosmology with $\Omega_{\rm m} = 0.3$, $\Omega_{\rm b} h^2
=0.02$ and $h=0.67$, where the current Hubble constant is $H_0 \equiv
h \times 100 \kms \mpc^{-1}$. The periodic volume has $7h^{-1}$
comoving Mpc on a side and $128^3$ initial resolution and has been
evolved to redshift 3. With nine refinement levels the maximum spatial
dynamic range is $6.5\tento{4}$, corresponding to 150 comoving parsec
maximum spatial resolution. The dark matter particles have masses of
$1.74\tento{7}\msun$, with masses of individual halos falling in the
range $1.5\tento{9}-10^{11}\msun$. The non-equilibrium chemistry of
six species, all ions of hydrogen and helium and free electrons, are
followed. Reionization in the computations was simulated using a
uniform background radiation field given by Haardt \& Madau (1996),
with a spectral slope of $\alpha=1.8$ starting at redshift 7.  Two
simulations were performed, both allowing for star formation but only
one accounting for strong stellar feedback. The two simulations
bracket the observed values of the comoving star formation rate at
high redshifts.

Star formation and stellar feedback are followed by a recipe similar
to the one presented by Cen \& Ostriker (1993), which has been
implemented and tested in detail by O'Shea et al.~(2002, in
preparation). Some of the relevant details of this algorithm are
summarized below.

\subsection{Star Formation and UV Emissivity }

When a contracting region that shows rapid cooling (its cooling
timescale $\tcool$ becomes shorter than the dynamical timescale
$\tdyn$) and is Jeans unstable is identified it is converted into a
collisionless stellar particle with a minimum mass of $10^6\msun$. The
mass $m_*$ of a stellar particle is recorded at the time of its
creation $t_{\rm form}$, and also we store $\tdyn$ derived from the
average density of the region at that moment. The star formation rate
(Fig.~\ref{fig:all1}) accounted for by this particle is spread over
several $\tdyn$ according to

\begin{equation}
{dM_{\rm SF}\over dt}(t)={m_*\over\tdyn}\left({t-t_{\rm form}\over\tdyn}\right)
\exp\left(-{t-t_{\rm form}\over\tdyn}\right),
\label{eq:sfrate}
\end{equation}

\noindent
with a corresponding UV luminosity of simply

\begin{equation}
L_{\rm UV}=\euv c^2{dM_{\rm SF}\over dt},
\label{eq:lumin}
\end{equation}

\noindent
where we have assumed a range $5\times 10^{-6}\le\euv\le 4\times
10^{-5}$. The justification for eq.~(\ref{eq:sfrate}--\ref{eq:lumin})
have been discussed in Cen \& Ostriker (1993). This range gives
reionization before $z\sim 6$ and yields moderate values of the UV
background after complete overlap, for the run with quenched star
formation due to strong stellar feedback. Note that this range is
slightly higher than the interval adopted by Gnedin (2000). Clearly,
$\euv$ depends on a number of hidden parameters, such as the initial
mass function (IMF), the stellar spectral energy distribution (SED)
and the escape fraction $\fesc$ of ionizing radiation from host
halos. At a grid resolution of $70\kpc$ comoving, we are unable to
accurately compute the internal absorption of UV radiation inside
galaxies. As such, our UV emission parameter $\euv$ must be viewed as
the amount of ionizing radiation escaping the galaxy. For comparison,
our range of chosen values of $\euv$ maps into a range of escape
fractions $0.07\le\fesc\le 0.55$ if we adopt the model of
\citet{ciardi00}.

\section{Radiative Transfer}

We solve the RT equation on a uniform Cartesian grid of $128^3$ cells
(i.e., the root grid of the AMR simulation). The total radiation field
is divided into direct photons coming from point sources and the
diffuse flux originating from recombination and bremsstrahlung
processes in the volume. From the computational perspective, the only
difference between these two components is the number of active
sources. For example, as discussed in details below, by the end of the
run at $z=4$ our entire volume features $\sim 2\times 10^3$ star
forming (SF) cells and $\sim 2\times 10^6$ gas cells. The light
crossing time across the computational volume is much shorter than the
timescales for opacity changes. This allows us to omit the
time-dependent term in the RT equation (see also Abel et al.~1999 and
Norman, Paschos, \& Abel 1998). The only exception is a logical switch
which does not allow any significant changes in the luminosity
(greater than $10\%$) of individual stellar sources to propagate
faster than the speed of light. At each timestep we compute the
luminosity $L_{\rm UV}^n$ of every source and compare it to its
luminosity $L_{\rm UV}^{n-1}$ at the previous timestep. If it changes
by more than $10\%$ then only those cells which are within $c\Delta
t^n$ from the source will be affected, and as we cross the $c\Delta
t^n$ radius along each ray, we multiply the photon count by $L_{\rm
UV}^{n-1}/L_{\rm UV}^n$.  We include this switch because the box light
crossing time $\tau_{\rm box}$ (3--6 Myr) is not negligible compared
to the timestep (1 Myr), although in practice it is not always
important because the luminosity of individual sources rises and
declines slowly on $\tdyn$ (Myr to tens of Myr) which is large
compared to $\tau_{\rm box}$.

The radiation field is reconstructed from the 3D source function at
each timestep. In this simulation, for both the direct and diffuse
components, we adopt a photon conserving scheme similar to the one
suggested by Abel et al.~(1999). In our present model, for both
components we assume a power law (with index $\alpha=5$) spectrum of
the photon number density. All radiative transfer is currently done in
a single frequency group above the hydrogen Lyman limit $\nu_{\rm
Ly}$.


\subsection{Direct Stellar Photons}

Transfer of direct stellar photons is computed explicitly using the
quad-tree technique of \citet{abel01}. We refer the reader to this
paper for the details, here we just briefly outline the
algorithm. Around each source we build a tree of rays, starting with
12 rays at level $l=1$, with individual ray segments splitting
recursively into four child rays each as they move further away from
the source. The total number of rays at any given level $l=1,2,...$ of
the hierarchy is $12\times 4^{l-1}$, with each ray corresponding to
the same equal area of the sphere. The length of individual ray
segments is chosen such that on one hand each cell in the volume is
connected to the stellar source, and at the same time one keeps the
number of ray segments to a minimum. At the top of this hierarchy at
$l=0$ (at the location of the stellar particle) we inject photons
which we then propagate along the tree. As a ray segment crosses a
given grid cell, we compute the rates of individual photochemical
reactions in that cell and update the photon count along the
corresponding ray segment. Each ray segment splits recursively into
four individual sub-segments, until either we cross the simulation
volume twice (we are assuming periodic boundary conditions), or the
photon flux is attenuated below $10^{-10}$ of its original value.

Sources of stellar radiation are identified in the hydro simulation,
using converging flow, cooling time and Jeans mass criteria, as
discussed above. The number of stellar sources rises sharply with
decreasing redshift, reaching $76,837$ by $z=5$ for the run with
stellar feedback.  We group these into $1,822$ cells at that time, at
$128^3$ resolution, and we perform exact radiative transfer with
quad-trees around each of these sources. Once most of the box is
transparent to ionizing photons, one has to construct the complete
hierarchy of trees, resulting in $12\times 4^{l-1}$ ray segments in
the outer ($l=10$) layer.  We have to build this tree around each
source, and this noticeably slows down the entire calculation.  But by
grouping nearby sources together once their ionization fronts overlap,
the calculation can be sped up approximately a factor of ten.

\begin{figure*}
\plotone{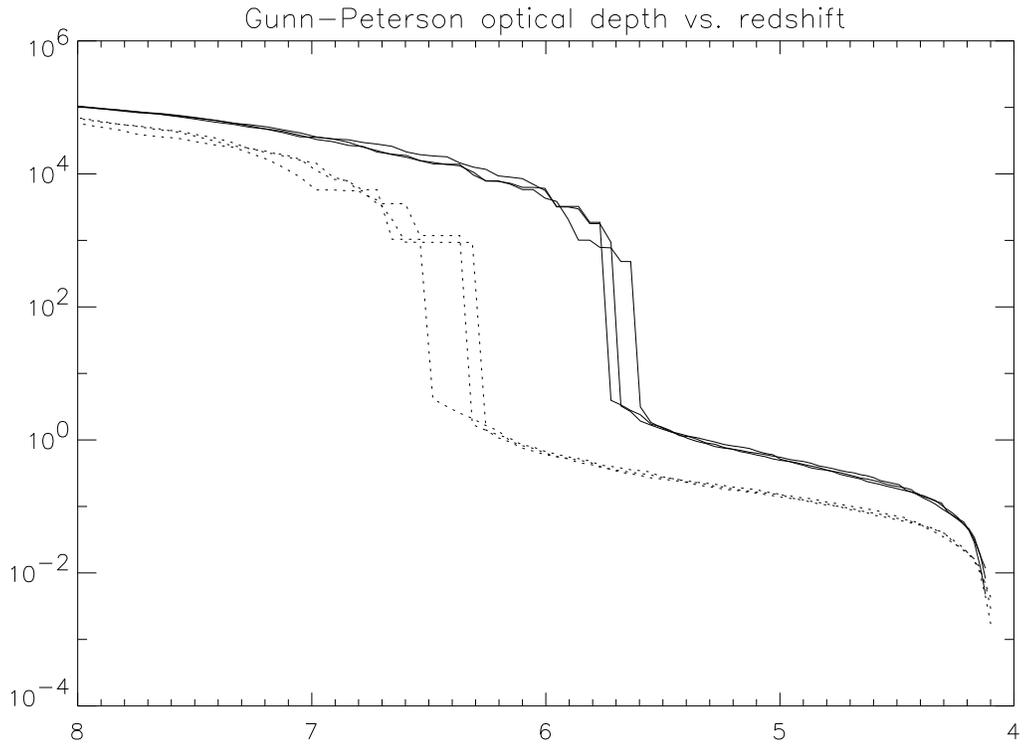}
\caption{Gunn-Peterson optical depth for models with stellar feedback
and $\euv=5\times 10^{-6}$ (solid lines) and $\euv=8\times 10^{-6}$
(dotted lines), along three random lines of sight.
\label{fig:lyalpha_3}}
\end{figure*}

If two emitting cells are in immediate proximity of each other, then
the cumulative effect on the IGM at some distance from these sources
will be exactly equal to the sum of their individual contributions.
One can treat these cells as a single point source, given that the IGM
is transparent to ionizing photons within the radius of
interest. Close to those sources one still might need to treat them
separately, especially when the sources are just starting to ionize
the surrounding medium. However, we find that once the medium close to
SF regions is in a highly ionized state, we obtain virtually identical
results for the ionizational state and temperature of the gas through
the entire \hii region when we group neighboring point sources. A
simple explanation for this result is that in either case we
explicitly conserve photons and have the same balance of the number of
ionizations to the number of recombinations, once the UV background
establishes itself in the common \hii bubble. Stellar sources are
highly clustered into few hundred halos in our $10\mpc$ volume (marked
by red dots in Fig.~\ref{fig:mosaic_oka}--\ref{fig:mosaic_oki}). In
addition, diffuse radiation is also almost the same in both cases,
since the temperature of the \hii region behind the I-front is nearly
constant (Abel \& Haehnelt 1999), independent of the distance to the
ionizing source, while the diffuse emissivity depends mostly on the
local density and not the location relative to the proto-galaxy.

To group sources together, we borrow a bottom-up tree algorithm from
McMillan \& Aarseth (1993). The procedure for this binning is
straightforward. At each timestep, we go through all point sources in
the volume and find the closest pair between which the optical depth
does not exceed $\tau_{\rm min}$. If this separation is also smaller
than some threshold value $d_{\rm min}$, then we merge these two
sources into a new source with the location weighted by the two
luminosities, and repeat the procedure of finding the closest pair
with $\tau<\tau_{\rm min}$. This technique will be reasonably accurate
providing that $\tau_{\rm min}$ is small enough, since in almost all
cases a negligible optical depth between two sources means that each
source blows a significant \hii region around them.

Note that unreasonably large values of $\tau_{\rm min}$ will bin two
sources which have not had enough time to create extended \hii
regions. Then, after these points are merged, the photon production
site suddenly moves further from these two partially ionized regions,
and now there are not enough ionizing photons to balance
recombinations; as a result, the neutral fraction goes up, and at the
next timestep the optical depth between the two sources jumps. Since
at every timestep the tree of sources is constructed from scratch,
these two sources will not be binned until they again ionize their
immediate surroundings. This oscillatory behavior is a clear sign of
unphysical binning and, hence, can be used as a test to choose the
optimal value of $\tau_{\rm min}$.  As a compromise between accuracy
and speed we chose $d_{\rm min}=1.16h^{-1}\mpc$ and $\tau_{\rm
min}=0.1$.

\subsection{Diffuse Radiation}

The diffuse component coming from recombinations and bremsstrahlung in the
gas and from the background UV flux are computed separately using a network of
long rays crossing the entire volume. At each timestep, we calculate the
emissivity of each cell and redistribute the corresponding number of photons
equally among all rays crossing this cell. As we proceed along each ray, we add
diffuse photons and compute sinks due to the local opacity, conserving
photons explicitly.

By connecting each possible pair of boundary cells, one could ensure
that every pair of cells inside the volume is connected.  However, for
a resolution of $N$ grid points in every direction, that would require
one to compute photon conservation along $15\times N^4$ rays.  We
refer to this as the \emph{full angular resolution}, and it is a
prohibitively large number for $N\gt 128$. Fortunately, the diffuse
intensity does not exhibit as strong angular and spatial variation as
direct photons from ionizing sources. Numerical tests of the sort
described in \citet{razoumov99} show that for the typical amount of
clumping found in cosmological simulations it would suffice to use
$\sim 10^{-3}$ of the number of rays one would employ with the full
angular resolution, simply because radiating regions occupy a
significant fraction of the volume. For example, in our $128^3$
simulations we get satisfactory results with $\sim 4\times 10^6$ rays
sampling the diffuse radiation.

Emissivity from radiative recombinations and bremsstrahlung are
computed using atomic models for hydrogenic and He-like ions. For
hydrogenic ions the frequency-dependent recombination and free-free
emissivity is (Hummer 1994)

\begin{equation}
\label{eq:emissivity}
\epsilon_\nu={\kb\te\over 4\pi}\nh n_{\rm ion}x_{\rm e} \left[
\sum_{n=1}^\infty \phi_n(\te,Z,\nu)+\phi_{\rm ff}(\te,Z,\nu)\right],
\end{equation}

\noindent
where

\begin{equation}
\phi_n(T_e,Z,\nu)={c\alpha^3\over\sqrt{\pi}}Z^3\lambda^{5/2}n^{-2}{\cal E}
(1+n^2{\cal E})^2e^{-\lambda{\cal E}}\cdot{\sigma_n(Z,{\cal E})\over
Z^2\nu_1},
\end{equation}

\begin{equation}
{\cal E}={\nu\over Z^2\nu_1},
~~~~\lambda={Z^2\hp\nu_1\over\kb\te},
\end{equation}

\begin{equation}
\phi_{\rm ff}(T_e,Z,\nu)={8\over 3\pi^2}\left(\pi\over 3\right)^{1/2}\alpha^2
 \lambda_{\rm c}^2cZ\lambda^{1/2}e^{-u}
 g_{\rm ff}(u,\lambda){\hp\over\kb\te}.
\end{equation}

\noindent
Here $\alpha$ is the fine-structure constant, $\nu_1$ is the hydrogen
Lyman limit, $Z$ is the atomic number, $\kb$ and $\hp$ are the
Boltzmann and Planck constants, and $\sigma_n(Z,{\cal E})$ is the
cross-section of photoionization for level $n$. Similarly, for the
free-free coefficient we have introduced the free-free Gaunt factor
$g_{\rm ff}(u,\lambda)$, the Compton wavelength $\lambda_{\rm c}$, and
$u\equiv h\nu/\kb\te$ (Hummer 1994).  The quantity $n_{\rm ion}$ is
either $n_{{\rm H}^+}$ for hydrogen or $n_{{\rm He}^{++}}$ for doubly
ionized helium, while $x_{\rm e}$ is the fraction of free electrons
relative to hydrogen.  We take the numerical values for the hydrogenic
photoionization cross-sections $\sigma_n(Z,{\cal E})$ from Storey \&
Hummer (1991) and the free-free Gaunt factor $g_{\rm ff}(u,\lambda)$
from Hummer (1988), and integrate eq.~(\ref{eq:emissivity})
numerically to get

\[
\epsilon=\int_{\nu_{\rm Ly}}^\infty\epsilon_\nu d\nu
\]

\noindent
for our single frequency group. The expression for emissivities due to
recombinations to the $nlS$ state of neutral helium is similar and can
be found in Hummer \& Storey (1998). In our calculations we neglect
the emissivity from di-electronic recombinations of helium.

Note that the temperature of the photoionized gas does not rise above
a few $\times10^4$K (Abel \& Haehnelt 1999).  Hence, recombination and
bremsstrahlung photons alone will contribute little to the radiation
field above the Lyman limit and, therefore, have relatively low effect
on the shape and the propagation speed of I-fronts. Of course, this
does not apply to the UV background itself, which has a profound
effect after the overlap of individual \hii regions.

\subsection{Boundary Conditions}


Direct unabsorbed photons from stellar sources inside the box are
traced for two full lengths of the computational volume without the
cosmological terms, assuming simple periodic boundary conditions. Once
these photons reach the boundary after two lengths, they are added to
the global pool of background photons. On the other hand, all diffuse
(recombination, bremsstrahlung and background) photons which are
already present in the volume are followed just for one full length of
the box, and after crossing the boundary they are added to the pool of
background photons. Thus at every timestep photons which have just
been added to the background come from two contributions. We then
apply the cosmological effects -- dilution of radiation due to cosmic
expansion and the Doppler shift -- to all background photons, and
using the diffuse solver we inject these photons uniformly into the
photoionized part of the IGM which is already exposed to the average
UV background. By construction, these are the cells for which the
optical depth from the boundary (averaged over all directions) is less
than unity.

\subsection{Chemistry}

Parallel to our RT calculation, we compute the non-equilibrium
chemistry of nine species: ${\rm H}$, ${\rm H}^+$, ${\rm He}$, ${\rm
He}^+$, ${\rm He}^{++}$, ${\rm H}^-$, ${\rm H}_2^+$, ${\rm H}_2$, and
${\rm e}^-$ (Abel et al.~1997) using the numerical algorithm described
in \citet{anninos97}. As we do raytracing for both the direct and the
diffuse components, at each grid cell we store all photo-chemical
reaction rates, which are then used to update ionization
fractions. Similar to \citet{sokasian01}, for recombination rates we
use the clumping factor extracted from the input simulations sampled
at $512^3$ resolution (i.e., to two levels of refinement in the AMR
grid hierarchy), however, unlike in their paper, we do not assume
photoionization equilibrium solving the full rate equations
instead. The temperature of the medium is computed explicitly at each
grid cell from energy conservation taking into account heating due to
photoionizations and cooling due to radiative recombinations and
bremsstrahlung, collisional excitations and collisional ionizations,
molecular hydrogen cooling, as well as Compton cooling. In practice,
since we do not solve hydrodynamical and radiative transfer equations
simultaneously, the evolution of species other than neutral and
ionized hydrogen -- such as molecular hydrogen, for example -- is not
crucial in this model, although these species were formally included
in our calculations.

\subsection{Summary of Approximations}

Let us briefly summarize the approximations we take. First of all,
hydro and RT are computed separately, i.e.~photoheating due to
diffuse radiation has no dynamical effect on the gas.  Secondly, we
omit the time derivative in the RT equation in favor of simple photon
statistics, so that at this stage we cannot properly compute ${\cal
O}(v/c)$ effects. We also employ finite angular resolution for the
diffuse flux, but fortunately, due to the almost isotropic nature of
this diffuse component and its small value, this approximation has
negligible effect on the shape and speed of I-fronts.  In addition, we
group nearby stellar sources once they create significant \hii regions
around them.  Finally, we use just one spectral group for the
radiation, therefore, we cannot study He reionization in these
particular calculations.

\section{Results}

\begin{figure*}
\plotone{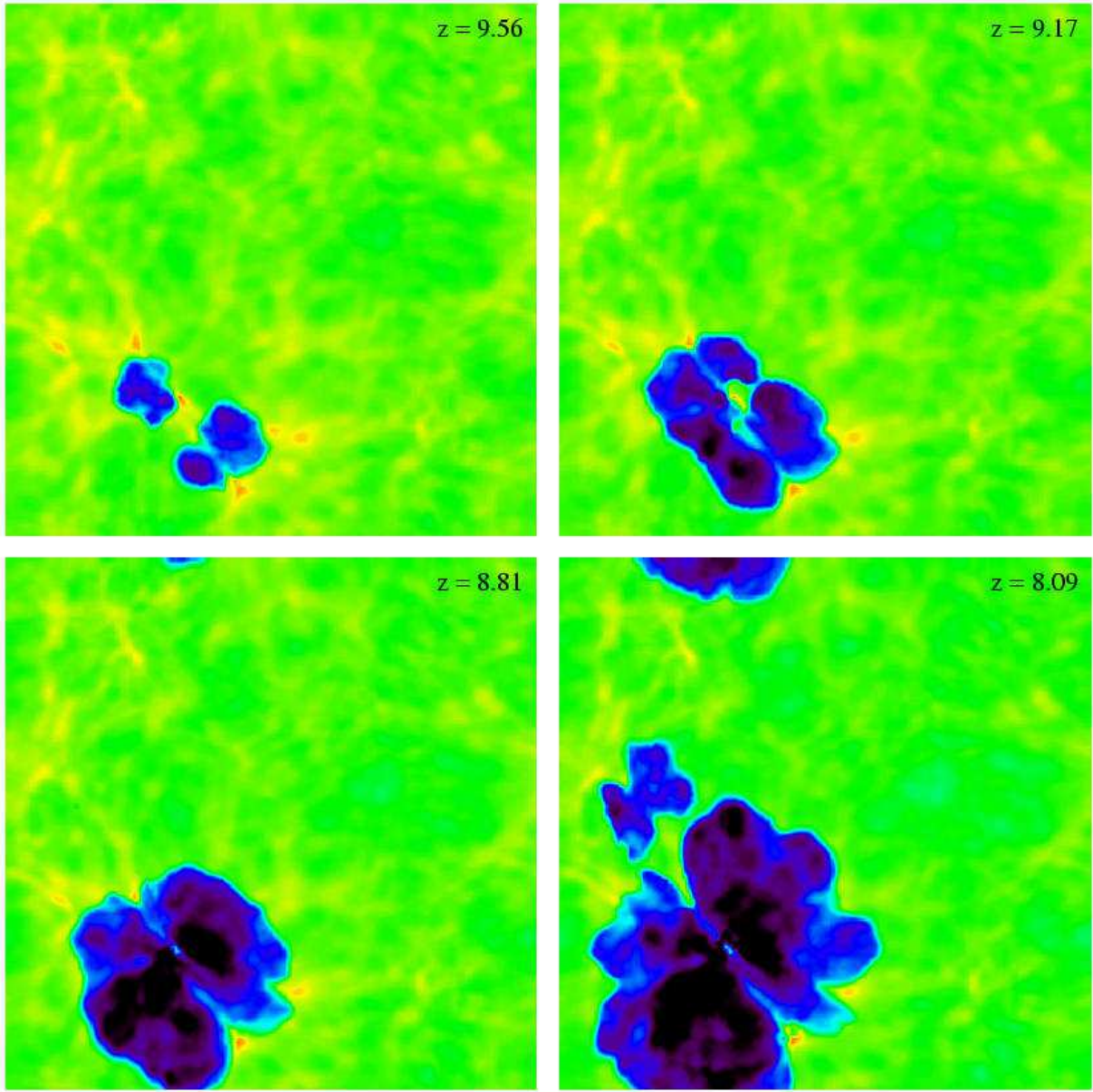}
\caption{\hi density distribution in a thin slice at four
output redshifts for the $\euv=5\times 10^{-6}$ model with stellar feedback,
showing how ionizing radiation breaks out into the voids.
\label{fig:2d_oka}}
\end{figure*}

One question which can be immediately addressed by our simulations is
whether underdense regions (like voids) are ionized before overdense
regions (like filaments).  Our full RT simulations in a $7h^{-1}\mpc$
volume show a picture of stellar reionization in which photons
initially do not travel far from ionizing sources, in contrast with
images from the simulations by Gnedin (2000). Fig.~\ref{fig:2d_oka}
shows that although ionizing photons stream preferentially into the
voids, \hii regions do not grow bigger than a few hundred thousand kpc
until after most of the gas close to sources is already ionized. This
result differs from that of Gnedin (2000), who found voids ionizing
before \hii regions had percolated. This counter-intuitive result
stems from Gnedin's inclusion of a homogeneous UV background which is
not present in our simulation until the volume is fully ionized. In a
sense, radiative transfer in our work is much more local, since the
mean free path of ionizing photons in the fully ionized medium is much
smaller than the box size. Host halos -- and on larger scales dense
filamentary structures around star forming proto-galaxies -- serve as
efficient sinks of radiation delaying global reionization by a
significant fraction of the Hubble time. Only after these regions are
ionized does radiation break out into the low-density IGM, sweeping
quickly through the rest of the volume.

A related question -- on how many photons per hydrogen atom would be
needed to cause complete overlap -- has attracted a lot of attention
in recent literature (Miralda-Escud\'e, Haehnelt, \& Rees 2000, Gnedin
2000, Haiman, Abel, \& Madau 2001). In our simulations we see that
about one ionizing photon per hydrogen atom in the volume has been
produced by the time of overlap. However, as discussed in Gnedin
(2000), such a low value is the direct result of our low resolution
($55h^{-1}\kpc$ comoving, or $14h^{-1}\kpc$ for recombinations with
the $512^3$ clumping factor), and does not contradict the conclusion
of Haiman et al.~(2001) that recombinations in mini-halos (with radii
of $1\kpc$ and smaller) would raise the required number of ionizing
photons by an order of magnitude. Also note, that the assumed
efficiency parameter $\euv$ already contains $\fesc\sim 10\%$ if we
assume the Salpeter IMF, that is, a significant fraction of ionizing
photons never gets out of a resolution element. In the near future we
are planning to increase our resolution to explicitly compute a larger
part of the escape fraction.

\begin{figure*}
\plotone{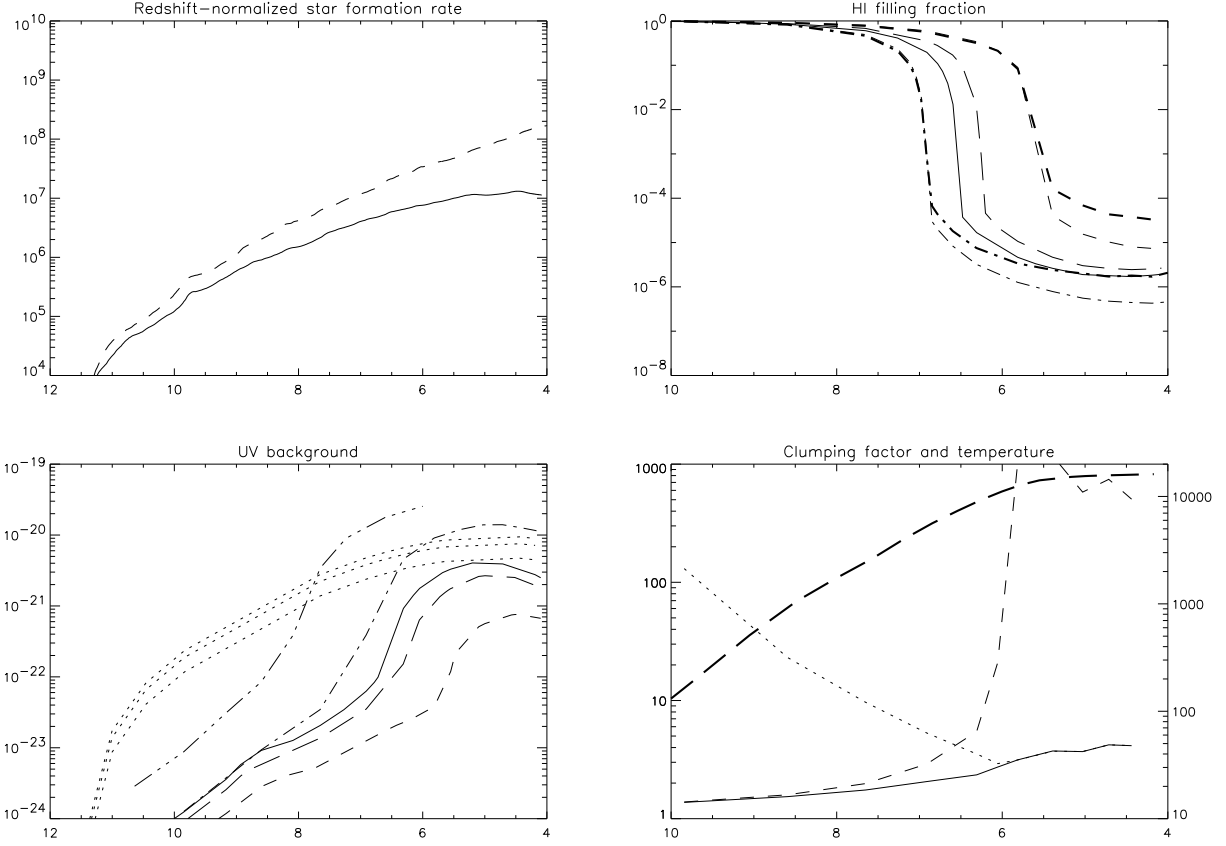}
\caption{Upper left panel: redshift-normalized average star formation
rate (in $\msun\mpc^{-3}$ per unit redshift interval), with (solid
line) and without (dashed line) stellar feedback. Lower left panel:
spatially-averaged Lyman-limit background intensity (in
$\erg\cm^{-2}\s^{-1}\hz^{-1}\ster^{-1}$) as a function of redshift for
models with $\euv=5\times 10^{-6}$ (short-dashed line), $8\times
10^{-6}$ (long-dashed line), $10^{-5}$ (solid line), and $4\times
10^{-5}$ (dash-double-dotted line) with stellar feedback, and the
dash-dotted line shows the model with $\euv=5\times 10^{-6}$ and no
stellar feedback. The three dotted lines show the integrated UV
background if there were no absorbing medium in the volume, for the
same amount of star formation at $\euv=10^{-5}$, $8\times 10^{-6}$ and
$5\times 10^{-6}$ from top to bottom, respectively. Upper right panel:
volume-weighted (four thin lines) and mass-weighted (two thick lines)
\hi filling fraction, for models with stellar feedback and
$\euv=10^{-5}$ (solid line), $8\times 10^{-6}$ (long-dashed line),
$5\times 10^{-6}$ (short-dashed line), and without stellar feedback
and $\euv=5\times 10^{-6}$ (dash-dotted line). Lower right panel:
total (solid line), \hi (short-dashed line) and \hii (dotted line)
clumping factors, and volume-weighted temperature (thick long-dashed
line), for $\euv=5\times 10^{-6}$ with stellar feedback.
\label{fig:all1}}
\end{figure*}

\begin{figure*}
\epsscale{0.9}
\plotone{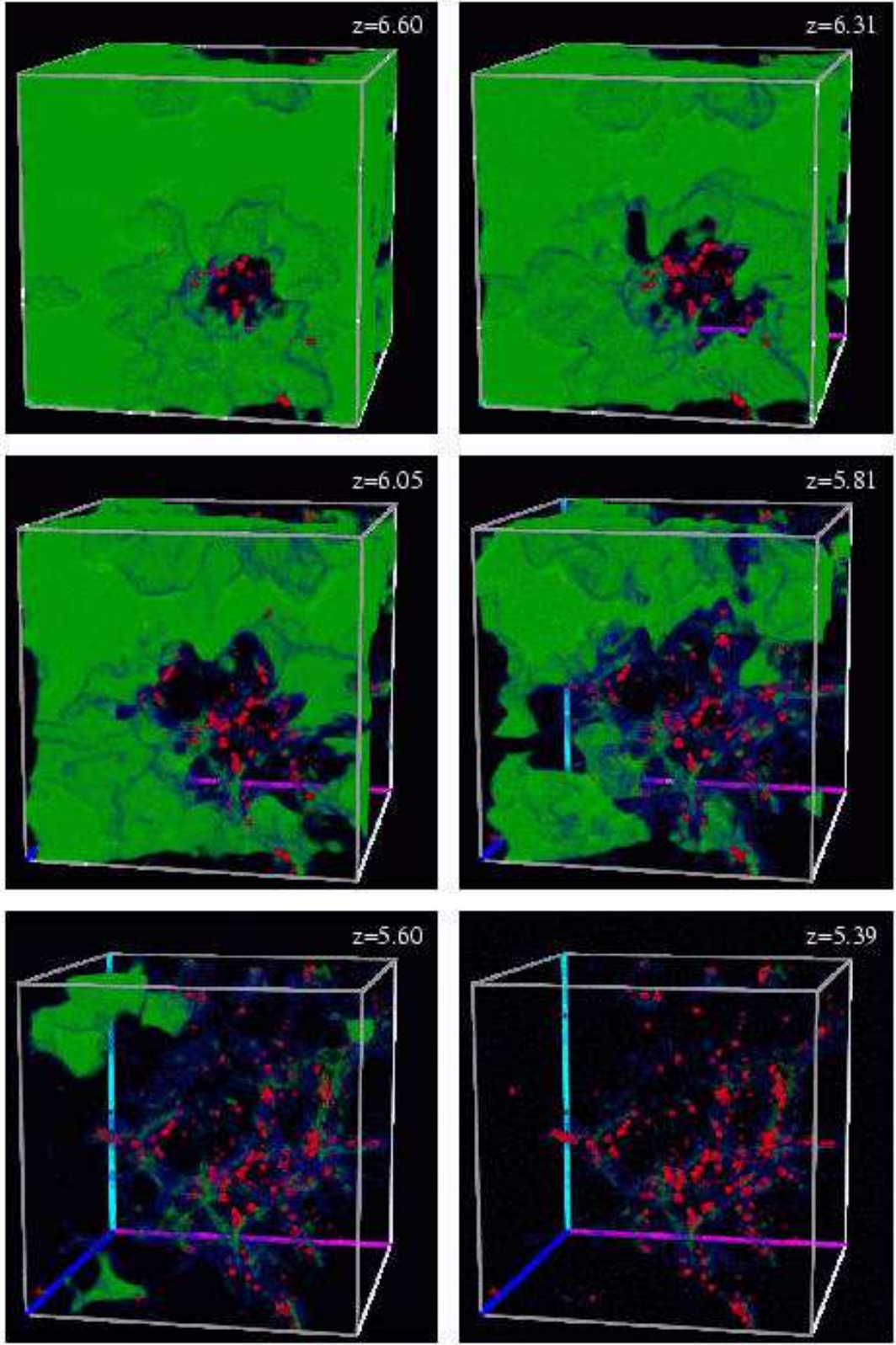}
\caption{Volume visualization of the \hi density distribution (green to
blue) and star forming protogalaxies (red) at six output redshifts
for the $\euv=5\times 10^{-6}$ model with stellar feedback. The first panel
corresponds to the time when ionizing photons from most massive halos
had already cleared their way to the nearest voids.
\label{fig:mosaic_oka}}
\end{figure*}

\begin{figure*}
\epsscale{0.9}
\plotone{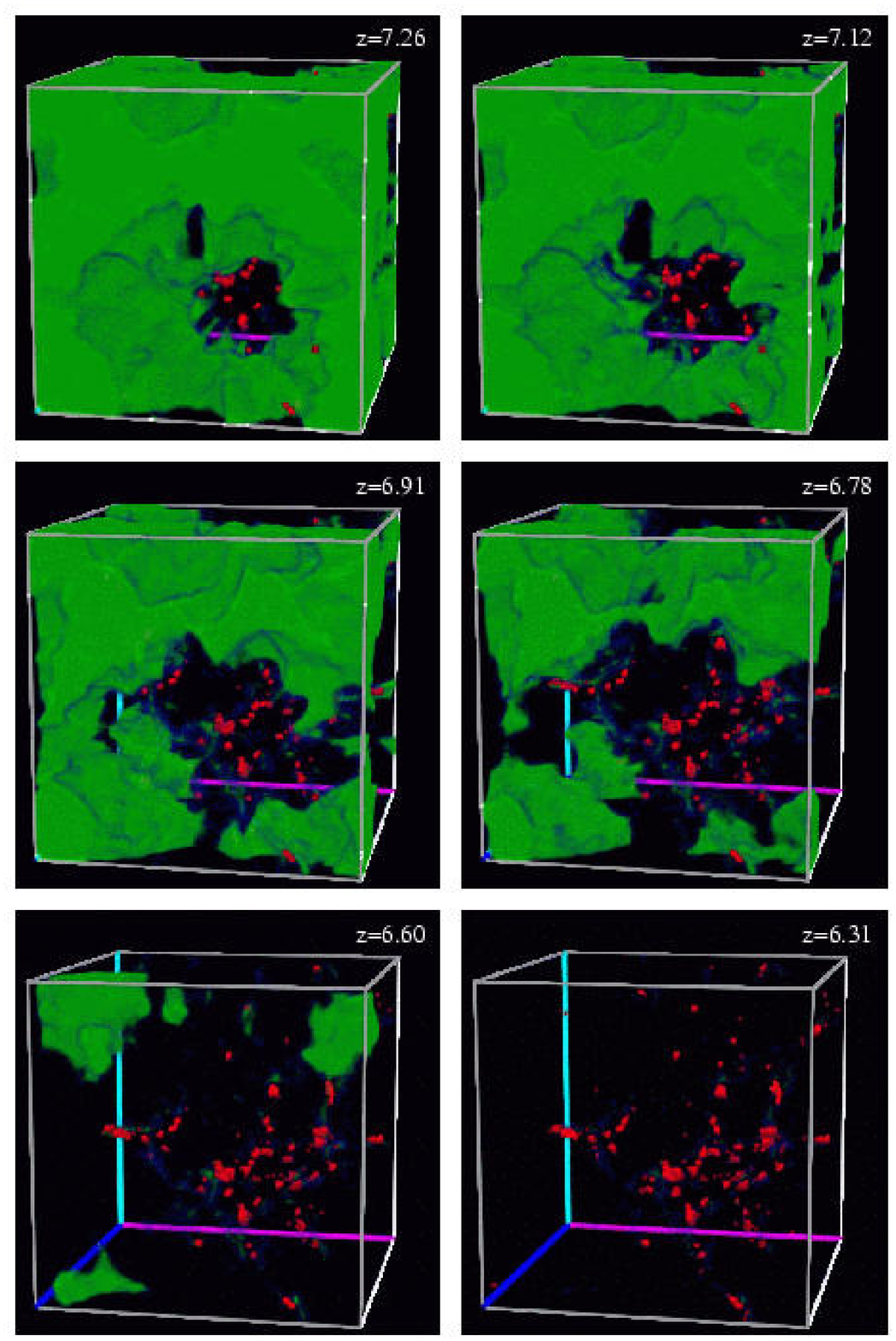}
\caption{Same as Fig.~\ref{fig:mosaic_oka} but for the
$\euv=10^{-5}$ model with stellar feedback.
\label{fig:mosaic_oki}}
\end{figure*}

The two phases -- initial escape and ionization of voids -- are
clearly separated in redshift (Fig.~\ref{fig:2d_oka} and
Fig.~\ref{fig:mosaic_oka}, respectively, both showing the same
model). These are the `pre-overlap' and the `overlap' in Gnedin's
terminology, and despite the differences in the details of how photons
are being deposited into the IGM in his and our models, we observe
similar stages of cosmic reionization. Figs.~\ref{fig:mosaic_oka} and
\ref{fig:mosaic_oki} show simultaneously the 3D neutral hydrogen
density distribution (green to blue in order of decreasing column
density) and the location of star forming protogalaxies (red) at six
output redshifts for two different UV production rates. These plots
confirm that the overlap is indeed fast compared to the initial escape
stage. In general, we find that the UV production efficiency in the
range $\euv=5\times 10^{-6}-4\times 10^{-5}$ leads to complete overlap
of \hii regions in the redshift interval $z=5.8-7.2$
(Fig.~\ref{fig:all1}). Lower values of $\euv$ would not give full
reionization until redshifts below those of the highest $z$ objects
already known, whereas a higher luminosity output would produce too
strong a UV background. The temperature of the ionized part of the IGM
stays virtually constant at an average of $T\sim 16,000{\rm K}$ in the
voids (see the volume-weighted temperature in the last panel in
Fig.~\ref{fig:all1}), rising to $20,000-40,000{\rm K}$ inside fully
ionized halos, and remaining below $8,000{\rm K}$ in the neutral
patches.

In Fig.~\ref{fig:lyalpha_3} we plot the Gunn-Peterson optical depth
(Fan et al.~2001)

\begin{equation}
\tau_{\rm GP}(z)=1.5\times 10^4 h^{-1}\Omega_{\rm m}^{-1/2}
\left(\Omega_{\rm b}h^2\over 0.02\right)
\int_0^z
\left(1+z'\right)^{1/2}
x_{\rm HI}(z')dz'
\end{equation}

\noindent
as a function of redshift, for two values of $\euv$ along three random
lines of sight. For the assumed star formation history, the range
$\euv=7$--$8\times 10^{-6}$ describes the observed statistics of very
high-$z$ Ly$\alpha$ absorbers (Becker et al.~2001). The late stages of
overlap already cause complete blanketing of the transmitted flux
blue-ward of the redshifted Ly$\alpha$ wavelength.  However, this
transition in the optical depth from $\tau_{\rm GP}\sim 10^3$ to
$\tau_{\rm GP}\sim{\rm few}$ demonstrates a scatter of $\Delta z\sim
0.2$, between different lines of sight.

After complete overlap and the rise of the UV background, the latter
comes to an equilibrium with the remaining neutral patches. All of the
low density gas, as well as dense filamentary regions in the vicinity
of star forming protogalaxies, become fully ionized.  However, a lot
of the filamentary structures remain largely neutral in lower
redshifts panels of Fig.~\ref{fig:mosaic_oka}. This is the slow
`post-overlap' stage observed by Gnedin (2000). The volume-weighted
\hi filling fraction stops falling rapidly once it reaches a plateau
at $10^{-5}$--$10^{-6}$, shortly after complete overlap, with the
exact number depending on the model. Ly$\alpha$ clouds are still being
ionized at lower redshifts, although in two of our models with ongoing
gravitational collapse recombinations are starting to dominate overall
in the volume around $z=4$. Note, however, that the final neutral
hydrogen fraction, as well as the UV background value would also be
sensitive to the size of the box.  This is because our volume is not
sufficiently large to have proper statistics of damped Lyman limit
systems, and, as a result, we somewhat overestimate the value of the
background field.

\section{Discussion}

Our current simulations demonstrate the feasibility of a full
radiative transfer calculation, in which a constantly evolving density
field and a source function could be computed self-consistently within
a hydro code. In a forthcoming paper, we will couple our radiative
transfer method with the {\sl enzo} (Bryan \& Norman 1999) hydro
code. The algorithm developed in this paper extends the achievable
resolution limit by combining both ray segments and sources into
hierarchical tree structures as we move farther from host
proto-galaxies. A further step would be to make the scheme adaptive on
many levels. Developing fully adaptive mesh refinement RT simulations
is a long-term goal. In the meantime one could devise an array of
shortcuts in order to speed up RT calculations. For example, ray trees
can be build in such a way that they accurately capture narrow
I-fronts without the need for performing excessive calculations in the
rest of the volume. One possible solution is to keep photo-chemical
reaction rates constant, without computing full RT in already ionized
or yet fully neutral regions, updating these rates only every 10th or
100th timestep \citep{sokasian01}. Many such measures will be tried as
we make collective progress towards modeling the physics of radiation
as realistically as possible in cosmological simulations. We believe
that our current computations represent a significant step along this
path.

The goal of this paper has been to present self-consistent
inhomogeneous radiative transfer in a fairly large cosmological volume
filled with young star-forming galaxies and `dark' extended absorbers.
At this point we cannot directly compute the physical escape fraction
of ionizing photons from their host halos, both due to poorly
constrained physics at sub-galactic scales and also due to the present
lack of spatial resolution.  Instead we have to rely on parameterizing
the escape fraction within an effective UV efficiency factor.
However, the technique described here -- when combined with the power
of AMR within hydro simulations -- can bring us closer to a correct
quantitative description of the relevant galaxy formation processes,
both prior to and after the beginning of cosmic reionization. Ideally,
these models should be combined with the full frequency-dependent
physics of a code such as {\sl CUBA} (Haardt \& Madau 2001) to perform
exact matching between numerical star formation and the observed
quantities at lower redshifts, such as the UV background or directly
measured luminosity functions.

Recently obtained spectra of the highest redshift quasars (Becker et al.~2001,
Djorgovski et al.~2001) show tantalizing evidence for the end of the
reionization process occurring at $z\simeq6$.  It may therefore be possible
to obtain data in the near future which directly probes the reionization
process.  The redshift evolution and spatial topology of reionization may
be measurable using sufficiently many lines of sight.  Comparison with models
such as ours may allow connections to be made between the observational
diagnostics and physical inputs to the reionization process: IMFs, SF rates,
relative contributions of AGN, sizes of the first collapsing mini-halos, etc.
A suite of simulations along the lines of those described here will be
extremely helpful in fulfilling this goal.

\acknowledgments

This work was carried out under the auspices of the Grand Challenge
Cosmology Consortium which is funded by NSF grant
AST-9803137. Numerical simulations were performed using the SGI Origin
2000 systems at the National Center for Supercomputing Applications at
the University of Illinois, Urbana-Champaign. DS is supported by the
Natural Sciences and Research Council of Canada.

We gratefully acknowledge Konstantinos Tassis for providing us with
his unpublished galaxy formation simulation data.

\clearpage


\begin{thebibliography}{}

\bibitem[Abel et al. (1997)]{abel97} Abel, T., Anninos, P., Zhang, Y.,
	\& Norman, M. L. 1997, New Astronomy, 2, 181
\bibitem[Abel (2000)]{abel00}
	Abel, T. 2000, Revista Mexicana de Astronomia y Astrofisica
	Conference Series, 9, 300
\bibitem[Abel \& Haehnelt(1999)]{ah99}
	Abel, T., \& Haehnelt, M. G. 1999, \apj, 520, L13
\bibitem[Abel et al.(1999)]{abel99}
	Abel, T., Norman, M. L., \& Madau, P. 1999, ApJ, 523, 66
\bibitem[Abel \& Wandelt(2001)]{abel01} Abel, T., \& Wandelt, B. D. 2001,
	astro-ph/0111033
\bibitem[Anninos et al.(1997)]{anninos97} Anninos, P., Zhang, Y., Abel, T.,
	\& Norman, M. L. 1997, New Astronomy, 2, 209
\bibitem[Becker et al.(2001)]{becker01}
	Becker, R. H., et al. 2001, AJ, in press, astro-ph/0108097
\bibitem[Bryan \& Norman (1999)]{bryan99}
	Bryan G. L., \& Norman M. L. 1999, in IMA Vol. 117, Structured
	Adaptive Mesh Refinement (SAMR) Grid Methods, ed. S. B. Baden,
	N. P. Chrisochoides, D. Gannon, \& M. L. Norman
	(New York: Springer), 165
\bibitem[Cen \& Ostriker(1993)]{cen93}
	Cen, R., \& Ostriker, J. P. 1993, ApJ, 417, 404
\bibitem[Chiu \& Ostriker(2000)]{chiu00}  Chiu, W. A., \& Ostriker, J. P. 2000,
	ApJ, 534, 507
\bibitem[Ciardi et al.(2000)]{ciardi00} Ciardi, B., Ferrara, A.,
	Governato, F., \& Jenkins, A. 2000, MNRAS, 314, 611
\bibitem[Ciardi et al.(2001)]{ciardi01} Ciardi, B., Ferrara, A.,
	Marri, S., \& Raimondo, G. 2001, MNRAS, 324, 381 
\bibitem[Djorgovski et al.(2001)]{djorgovski01}
	Djorgovski, S. G., Castro, S. M., Stern, D., Mahaba, A. 2001, ApJ,
	in press, astro-ph/0108069
\bibitem[Fan et al.(2001)]{fan01} Fan, X. et al. 2001, AJ, submitted,
	astro-ph/0108063
\bibitem[Gnedin(2000)]{gnedin00} Gnedin, N. Y. 2000, \apj, 535, 530
\bibitem[Gnedin \& Abel(2001)]{gnedin01}
	Gnedin, N. Y., \& Abel, T. 2001, astro-ph/0106278
\bibitem[Gnedin \& Ostriker(1997)]{gnedin97}
	Gnedin, N. Y., \& Ostriker, J. P. 1997, ApJ, 486, 581
\bibitem[Haardt \& Madau(1996)]{haardt96}
	Haardt, F., \& Madau, P. 1996, ApJ, 461, 20
\bibitem[Haardt \& Madau(2001)]{haardt01} Haardt, F., \& Madau, P. 2001,
	astro-ph/0106018
\bibitem[Haiman et al.(2001)]{haiman01} Haiman, Z., Abel, T., \& Madau, P.
	2001, \apj, 551, 599
\bibitem[Hummer(1988)]{hummer88} Hummer, D. G. 1988, ApJ, 327, 477
\bibitem[Hummer(1994)]{hummer94} Hummer, D. G. 1994, MNRAS, 268, 109
\bibitem[Hummer \& Storey(1998)]{hummer98}
	Hummer, D. G., \& Storey, P. J. 1998, MNRAS, 297, 1073
\bibitem[Loeb \& Barkana(2001)]{loebar01}
	Loeb, A., \& Barkana, R. 2001, ARA\&A, 39, 19
\bibitem[McMillan \& Aarseth(1993)]{mcmillan93}
	McMillan, S. L. W., \& Aarseth, S. J. 1993, \apj, 414, 200
\bibitem[Miralda-Escud\'e, Haehnelt, \& Rees (2000)]{miralda00}
	Miralda-Escud\'e, J., Haehnelt, M., \& Rees, M. J. 2000, ApJ, 530, 1
\bibitem[Norman et al. (1998)]{norman98} Norman, M. L., Paschos, P.,
	\& Abel, T. 1998, Memorie della Soc. Astron. Italiana, 69, 455
\bibitem[Razoumov \& Scott(1999)]{razoumov99}
	Razoumov, A. O., \& Scott, D. 1999, \mnras, 309, 287
\bibitem[Seljak \& Zaldarriaga (1996)]{seljak96} Seljak, U., Zaldarriaga, M.
	1996, ApJ, 469, 437
\bibitem[Sokasian et al.(2001)]{sokasian01} Sokasian, A., Abel, T.,
	\& Hernquist, L. E. 2001, astro-ph/0105181
\bibitem[Storey \& Hummer(1991)]{stohum91}
	Storey, P. J., \& Hummer, D. G. 1991, Comput. Phys. Commun., 66, 129
\bibitem[Umemura, Nakamoto \& Susa(1998)]{umemura98}
	Umemura, M., Nakamoto, T., \& Susa, H. 1998,
	in Numerical Astrophysics 1998, ed. S. M. Miyama \& K. Shibata (Kluwer)
\bibitem[Valageas \& Silk(1999)]{valageas99}
	Valageas, P., \& Silk, J.  1999, A\&A, 347, 1

\end{thebibliography}
\end{document}